\documentclass[applemac]{aa}
\usepackage{inputenc}
\usepackage{natbib}
\usepackage{txfonts}
\usepackage{graphicx}
\usepackage{color}
\bibpunct{(}{)}{;}{a}{}{,}

\begin{document}

\title{On the stability of non-isothermal Bonnor-Ebert spheres. II. The effect of gas temperature on the stability}
\author{O. Sipil\"a\inst{1,2},
J. Harju\inst{1,2},
\and M. Juvela\inst{2}
}
\institute{Max-Planck-Institute for Extraterrestrial Physics (MPE), Giessenbachstr. 1, 85748 Garching, Germany\\
e-mail: \texttt{osipila@mpe.mpg.de}
\and{Department of Physics, PO Box 64, 00014 University of Helsinki, Finland}
}

\date{Received / Accepted}

\abstract
{}
{We investigate the stability of non-isothermal Bonnor-Ebert spheres in the context of a model that includes a self-consistent calculation of the gas temperature. In this way, we can discard the assumption of equality between the dust and gas temperatures, and study the stability as the gas temperature changes with the chemical evolution of the cooling species.}
{We use a gas-grain chemical model to calculate the chemical evolution. The model includes a time-dependent treatment of depletion onto grain surfaces, which strongly influences the gas temperature as the main coolant molecule CO depletes from the gas. The dust and gas temperatures are solved with radiative transfer calculations. For consistent comparison with previous work, we assume that the cores are deeply embedded in a larger external structure, corresponding to visual extinction $A_{\rm V}^{\rm ext} = 10$\,mag at the core edge. We also study the effect of lower values of $A_{\rm V}^{\rm ext}$.}
{We find that the critical non-dimensional radius $\xi_1$, determining the maximal density contrast between the core center and the outer boundary, derived here is similar to our previous work where we assumed $T_{\rm dust} = T_{\rm gas}$; the $\xi_1$ values lie below the isothermal critical value $\xi_0 \sim 6.45$, but the difference is less than $10\,\%$. We find that chemical evolution does not affect notably the stability condition of low-mass cores ($< 0.75\,M_{\odot}$) which have high average densities and a strong gas-grain thermal coupling. In contrast, for higher masses the decrease of cooling owing to CO depletion causes substantial temporal changes in the temperature and in the density profiles of the cores. In the mass range $1-2\,M_{\odot}$, $\xi_1$ decreases with chemical evolution, whereas above $3\,M_{\odot}$, $\xi_1$ instead increases with chemical evolution. We also find that decreasing $A_{\rm V}^{\rm ext}$ strongly increases the gas temperature especially when the gas is chemically old, and this causes $\xi_1$ to increase with respect to models with higher $A_{\rm V}^{\rm ext}$. However, the derived $\xi_1$ values are still close to $\xi_0$. The density contrast between the core center and edge derived here varies between 8 to 16 depending on core mass and the chemical age of the gas, compared to the constant value $\sim 14.1$ for the isothermal BES.}
{}


\authorrunning{O. Sipil\"a et al.}

\maketitle

\section{Introduction}

The Bonnor-Ebert sphere (\citealt{Bonnor56}, \citealt{Ebert55}; hereafter BES), i.e., an isothermal gas sphere in hydrostatic equilibrium, has been used succesfully to approximate the density structures of prestellar cores \citep{Bacmann00, Alves01, Kandori05, Marsh14}. However, the assumption of isothermicity is not valid generally \citep{Zucconi01, Ward-Thompson02, Pagani04, Crapsi07, Juvela11}. To accommodate for a radial temperature profile, a non-isothermal version of the BES (referred here to as a modified Bonnor-Ebert sphere; MBES) has been studied in the literature \citep{Evans01, Galli02, Keto05, Sipila11}.

In previous studies of the MBES, it is either assumed that the dust and gas temperatures are equal, or the gas temperature has been derived based on some standard abundances for the cooling species. However, both approaches are approximations and do not hold generally. In Sipil\"a et al.\,(\citeyear{Sipila11}; hereafter Paper~I), we studied the stability of MBESs that are deeply embedded in a larger external structure, e.g., a molecular cloud, corresponding to a high visual extinction $A_{\rm V}^{\rm ext} = 10$\,mag at the edges of the studied cores. We also assumed that $T_{\rm dust} = T_{\rm gas}$, which holds well for low-mass cores with high average densities, but is not valid at lower density where the collisional coupling between gas and dust is weak. In the present paper, we aim to generalize the analysis of Paper~I by including in the stability calculations a self-consistent determination of the gas temperature. This is accomplished by calculating chemical evolution in model cores with a comprehensive gas-grain chemical model \citep{Sipila12, Sipila13}, followed by a determination of the gas temperature with a radiative transfer model at different time steps, taking advantage of the time-dependent chemical abundances. In this way, we can study the stability condition not only as a function of the core mass but also as a function of ``chemical time'', which is here defined as the chemical age of the core since some initial state (see Sect.\,\ref{ss:stabdeterm}). We can determine whether the results of Paper~I are significantly affected when $T_{\rm dust} \neq T_{\rm gas}$, especially in high-mass cases when the cores have low average densities and consequently weak gas-grain thermal coupling.

The paper is organized as follows. In Sect.\,\ref{s:method}, we discuss our model calculations in detail. In Sect.\,\ref{s:results}, we present the results of our calculations, and discuss them in Sect.\,\ref{s:discussion}. We present our conclusions in Sect.\,\ref{s:conclusions}.

\section{Method}\label{s:method}

This section outlines how the calculations are carried out in practice. The chemical model is discussed here only briefly, and we refer the reader to \citet{Sipila12} and \citet{Sipila13} for a complete description of the model.

\subsection{The MBES}

In what follows, we discuss the basic properties of the MBES in a rather concise form; A more detailed discussion on the MBES can be found in Paper~I.

The MBES differs from the BES in that it is non-isothermal. Assuming the ideal gas equation of state and hydrostatic equilibrium, and that $T = T(r)$, the density distribution of the MBES is given by
\begin{equation}\label{eqrhodist}
\frac{1}{r^2} \frac{d}{dr} \left( \frac{r^2}{\rho} \left[ T \frac{d\rho}{dr} + \rho \frac{dT}{dr} \right] \right) = - \frac{4\pi Gm \rho}{k} \, .
\end{equation}
This equation can be transformed into non-dimensional form by making the substitutions
\begin{equation}\label{eqrho}
\rho = \frac{\lambda}{\tau(\xi)} {\rm e}^{-\psi(\xi)}
\end{equation}
\begin{equation}\label{eqr}
r = \beta^{1/2}\lambda^{-1/2}\xi \, .
\end{equation}
In the above, $\lambda$ is the central density of the core; $\tau(\xi) = T(\xi)/T_{\rm c}$, where $T_{\rm c}$ is the central temperature of the core; $\beta = k T_{\rm c} / 4\pi G m$, where $k$ and $m$ are the Boltzmann constant and the average molecular mass of the gas (assumed here equal to 2.33 amu), respectively; $\xi$ and $\psi(\xi)$ are dimensionless variables ($\xi$ represents a non-dimensional radius). With these substitutions, Eq.\,(\ref{eqrhodist}) transforms to the modified Lane-Emden equation
\begin{equation}\label{eqMLE}
\xi^{-2} \frac{d}{d\xi} \left[ \xi^2 \tau \frac{d\psi}{d\xi} \right] = \frac{1}{\tau} {\rm e}^{-\psi} \, .
\end{equation}
We impose the following boundary conditionsÊat the core center: $\psi = 0$, $d\psi / d\xi = 0$, $\tau = 1$ and $d\tau / d\xi = 0$. These boundary conditions ensure that $T = T_{\rm c}$ and $\rho = \lambda$ at $\xi = 0$, i.e., at the center of the core. To solve Eq.\,(\ref{eqMLE}), one also needs to supply a temperature profile, which has to be determined externally -- we discuss the temperature calculations in more detail below. Finally, the density profile of the MBES is given by substituting the solution function $\psi$ and the temperature profile into Eq.\,(\ref{eqrho}).

\subsection{Determining the stability of the MBES}\label{ss:stabdeterm}

The non-dimensional radius $\xi$ is a free parameter -- for a given core mass, there exist a series of core configurations corresponding to different values of $\xi$. In what follows, the non-dimensional radius of each core configuration will be represented by $\xi_{\rm out}$. It should be noted that the solution to Eq.\,(\ref{eqMLE}) is unique to each core configuration (see Paper~I), so that the solution is applicable in the interval $0 < \xi < \xi_{\rm out}$ for each value of~$\xi_{\rm out}$ (i.e., separately for each core configuration).

\begin{figure}[t]
   \centering
   \unitlength=1.0mm
  \begin{picture}(60,165)(0,4)
\color{black} 

\put(30, 161){\oval(74,6)}
\put(30, 161){\makebox(0,0){ Fix core mass $M$ and non-dimensional radius $\xi_{\rm out}$ } }

\put(30, 157){\vector(0,-4){2.5}}

\put(30, 151){\oval(96,6)}
\put(30, 151){\makebox(0,0){ Construct a BES corresponding to the chosen values of $M$ and $\xi_{\rm out}$ } }

\put(30, 147){\vector(0,-4){2.5}}

\put(30, 141){\oval(52,6)}
\put(30, 141){\makebox(0,0){ Calculate a dust temperature profile } }

\put(30, 137){\vector(0,-4){2.5}}

\put(30, 131){\oval(22,6)}
\put(30, 131){\makebox(0,0){ Solve Eq.\,(\ref{eqMLE}) } }

\put(30, 127){\vector(0,-4){2.5}}

\put(30, 121){\oval(55,6)}
\put(30, 121){\makebox(0,0){ Calculate central density from Eq.\,(\ref{eqBEmass}) } }

\put(30, 117){\vector(0,-4){2.5}}

\put(30, 111){\oval(55,6)}
\put(30, 111){\makebox(0,0){ Calculate density profile from Eq.\,(\ref{eqrho}) } }

\qbezier(2,111)(-20,127)(3,140)
\put(3.1,140){\vector(2,1){0.5}}
\put(0, 130){\makebox(0,0){ Iteration } }

\put(30, 107.5){\vector(0,-4){2.5}}

\put(30, 99){\oval(75,11)}
\put(30, 101){\makebox(0,0){ Divide the core into concentric shells and calculate } }
\put(30, 97){\makebox(0,0){ chemistry separately in each shell } }

\put(30, 93){\vector(0,-4){2.5}}

\put(30, 87){\oval(54,6)}
\put(30, 87){\makebox(0,0){ Calculate the gas temperature profile } }

\put(30, 83){\vector(0,-4){2.5}}

\put(30, 77){\oval(22,6)}
\put(30, 77){\makebox(0,0){ Solve Eq.\,(\ref{eqMLE}) } }

\put(30, 73){\vector(0,-4){2.5}}

\put(30, 67){\oval(55,6)}
\put(30, 67){\makebox(0,0){ Calculate central density from Eq.\,(\ref{eqBEmass}) } }

\put(30, 63){\vector(0,-4){2.5}}

\put(30, 57){\oval(55,6)}
\put(30, 57){\makebox(0,0){ Calculate density profile from Eq.\,(\ref{eqrho}) } }

\put(30, 53){\vector(0,-4){2.5}}

\put(30, 47){\oval(59,6)}
\put(30, 47){\makebox(0,0){ Calculate a new dust temperature profile } }

\qbezier(0,49)(-20,67)(-4,92)
\put(-3.9,92){\vector(2,3){0.5}}
\put(-3, 77){\makebox(0,0){ Iteration } }

\put(30, 43){\vector(0,-4){2.5}}

\put(30, 37){\oval(35,6)}
\put(30, 37){\makebox(0,0){ Extract $\beta$, $\lambda$ and $\psi(\xi_{\rm out})$ } }

\put(30, 33){\vector(0,-4){6.5}}

\put(30, 23){\oval(64,6)}
\put(30, 23){\makebox(0,0){ Repeat the above for different values of $\xi_{\rm out}$ } }

\put(30, 19.3){\vector(0,-4){6.5}}

\put(30, 9){\oval(74,6)}
\put(30, 9){\makebox(0,0){ Examine the stability of the solutions using Eq.\,(\ref{stability})  } }

\end{picture}

  \caption{Steps taken in deriving the critically stable configuration of an MBES of given mass. See text for detailed explanations of each step.}
      \label{fig:steps}
   \end{figure}
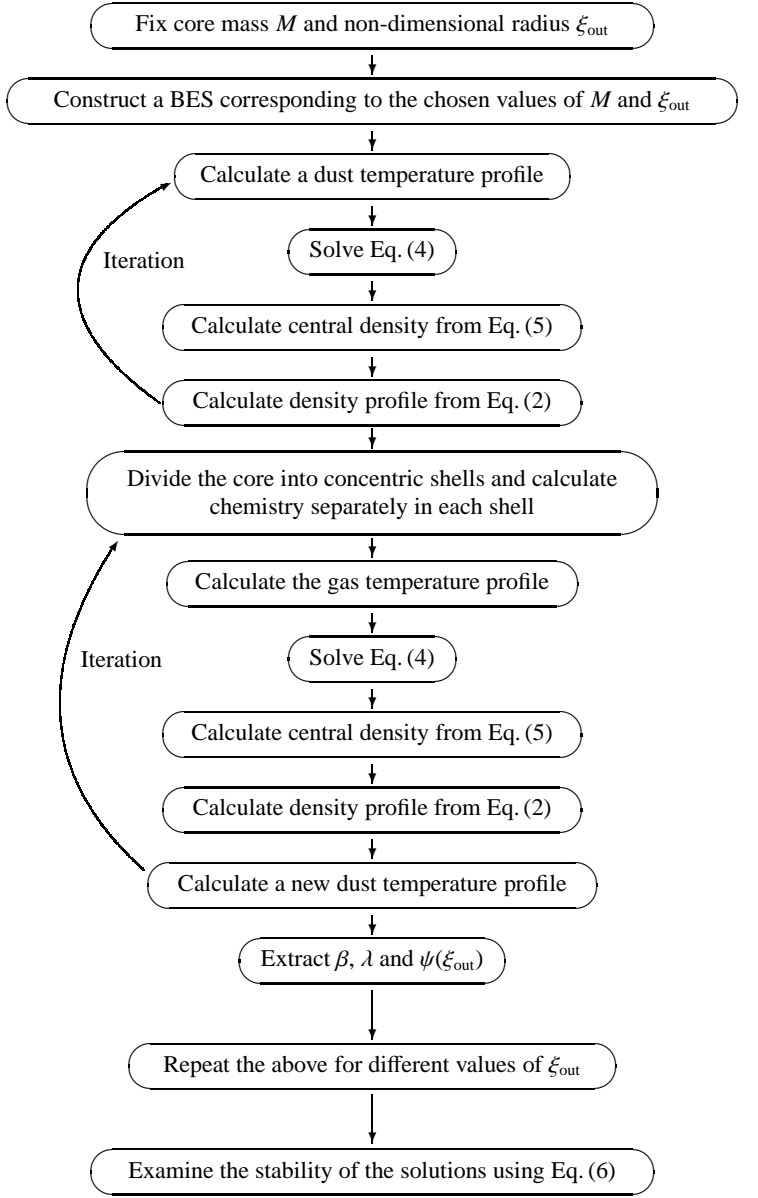

As discussed in Paper~I, determining the stability of an MBES of given mass is analogous to finding the core configuration which is critically stable. We construct the different core configurations in the same way as in Paper~I, i.e., by following an iterative process. To illustrate the process, outlined in Fig.\,\ref{fig:steps}, let us {\sl fix the core mass and the non-dimensional radius~$\xi_{\rm out}$}. We first construct a BES corresponding to the chosen values of core mass and $\xi_{\rm out}$, and adopting $T = 10\,\rm K$. After this step, we determine a dust temperature profile for the BES using radiative transfer modeling \citep{Juvela03, Juvela05}. The dust temperature profile is then used to solve Eq.\,(\ref{eqMLE}). In Paper~I, we considered two different dust models (from \citealt{OH94} and \citealt{LD01}), but here we only consider the former. This issue is discussed in Sect.\,\ref{ss:temp_eff}.

%

The solution to Eq.\,(\ref{eqMLE}) and the dust temperature profile are used to derive the central density of the MBES from the equation
\begin{equation}\label{eqBEmass}
M = 4\pi \, \beta^{3/2} \lambda^{-1/2} \xi_{\rm out}^2 \, \tau_{\rm out} \, \biggl({d\psi \over d\xi}\biggr)_{\rm out} \, ,
\end{equation}
where $\tau_{\rm out} = T(\xi_{\rm out})/T_{\rm c}$. After this step, the density profile of the MBES is calculated from Eq.\,(\ref{eqrho}). To ensure the consistency of the density profile with the temperature profile, we perform additional iterative calculations, calculating a new dust temperature profile and a new density profile in sequence. Both profiles typically reach their final solutions in a few iterations.

Next, we proceed to calculate the gas temperature. In Paper~I, we made the simplifying assumption that $T_{\rm dust} = T_{\rm gas}$. However, this is a rather crude estimate for cores with total hydrogen density $n_{\rm H} \lesssim 10^5\,\rm cm^3$, where the coupling between the dust and the gas is weak. In this paper, we carry out an updated analysis of the results of Paper~I, replacing the dust temperature with the gas temperature in all of the relevant formulae. The gas temperature is solved identically to the method discussed in \citet{Sipila12}. That is, we divide the first-approximation MBES density profile into concentric shells, solve the chemical evolution separately in each shell and extract chemical abundance profiles for the cooling species (see below) as functions of radial distance from the core center. The chemical model is adopted from \citet{Sipila13}; the model includes a full description of gas-grain chemistry and the deuterated forms of chemical species with up to 4 atoms.

The gas temperature is calculated with a Monte Carlo radiative transfer program \citep{Juvela97} which balances heating and cooling functions to solve the gas temperature. In the present model, the gas is heated externally by cosmic rays and by the photoelectric effect \citep{Goldsmith01}. However, the latter process is only important at $A_{\rm V} \lesssim 2$\,mag \citep{Juvela11}. Molecular line cooling is calculated for the following species: $\rm ^{12}CO$, $\rm ^{13}CO$, $\rm C^{18}O$, C, O and $\rm O_2$. Since the adopted chemical model does not include the isotopes of the various species, we adopt the isotopic ratios $\rm ^{12}CO / ^{13}CO = 60$ and $\rm ^{12}CO / C^{18}O~=~500$ (\citealt{Wilson94}). We also include the energy exchange (collisional coupling) between the gas and the dust grains following the description of \citet{Goldsmith01}\footnote{We note that \citet{Young04} have considered a stronger gas-grain coupling than in the model of \citet{Goldsmith01}, which might somewhat compensate for the depletion of coolant molecules at intermediate densities if adopted in our model.}. We implicitly assume that chemical timescales are longer than the radiative and dynamical timescales, so that a core is able to quickly reach a new equilibrium state as the chemistry evolves.

After the gas temperature profile has been calculated, it is used to solve Eq.\,(\ref{eqMLE}) again, producing a new density profile which is then used to redetermine successively the dust temperature, the chemical abundances and the gas temperature as outlined above. The iteration is repeated a few times until the density and the (gas and dust) temperature profiles converge toward their respective final solutions.

Carrying out the above iterative process for different values ofÊ$\xi_{\rm out}$, while keeping the core mass constant, we obtain $\beta$, $\lambda$ and $\psi$ as functions of $\xi_{\rm out}$ and use these to construct the pressure derivative
\begin{equation}\label{stability}
\frac{\delta p}{\delta V} = \frac{2 p}{3V} \frac{ \left[ \beta^{-1} \delta\beta + \lambda^{-1} \delta\lambda - \left( \frac{d\psi}{d\xi}\right)_{\rm out} \delta\xi_{\rm out} \right] } { \left[ \beta^{-1} \delta\beta -  \lambda^{-1} \delta\lambda + 2 \, \xi_{\rm out}^{-1} \, \delta\xi_{\rm out} \right] }  \, ;
\end{equation}
see Paper~I for details on how this expression is derived. Finally, the critically stable configuration corresponds to the lowest value of $\xi_{\rm out}$ for which the pressure derivative is zero -- all values of $\xi_{\rm out}$ above this value correspond to unstable configurations (\citealt{Bonnor56}; Paper~I). In the next section, we present the critical $\xi_{\rm out}$ values for a range of core masses derived as outlined above, and we label the critical $\xi_{\rm out}$ value for each core mass as $\xi_1$. Similarly, we label the values of $\beta$, $\lambda$ and $\psi(\xi_{\rm out})$ corresponding to critical cores as $\beta_1$, $\lambda_1$ and $\psi_1$, respectively.

Chemical abundances vary with time, and this influences the gas temperature as 1) atomic species are processed into molecules and 2) the main coolant species CO in its various isotopic forms freezes onto the dust grains. To investigate the effect of varying chemical abundances on the stability of the MBES, we have carried out the above analysis for four different time steps, corresponding to $5 \times 10^4$, $1 \times 10^5$, $5 \times 10^5$ or $1 \times 10^6$ years of chemical evolution since the initial state of the core (see below). In practice, we calculate $\xi_1$ for each core mass four times, extracting the chemical abundances at either $5 \times 10^4$, $1 \times 10^5$, $5 \times 10^5$ or $1 \times 10^6$ years. The different gas temperature profiles at the four time steps translate to marked differences in density profiles, and also in the critical $\xi_1$ values. This issue is discussed in the next section. We note that we assume external visual extinction of $A_{\rm V}^{\rm ext} = 10$\,mag in the chemical and radiative transfer calculations, so that the model cores are assumed to exist deeply embedded inside larger structures, such as molecular clouds. This choice ensures the possibility of direct comparison of our results with those of Paper~I.

In the chemical calculations, we assume that the gas is initially atomic, with the exception of hydrogen which is in molecular form \citep{Sipila12}. The choice of the chemical composition of the gas in the beginning of the calculation (labeled as $t_0$ in the figures) is a parameter of the model. The four timesteps defined above measure the extent of chemical development with respect to the atomic initial state. We note that an atomic chemical composition may not be consistent with an initial physical structure corresponding to a BES (as is assumed here) because chemical processing of the gas is expected to take place during the formation of the BES itself. However, here we do not attempt to impose constraints on ``absolute'' core ages, but to investigate if chemical evolution can influence the stability condition. Therefore, the chosen four timesteps are simply representative points along the chemical development track of a given core and its surroundings.

\section{Results}\label{s:results}

\begin{figure*}
\centering
\includegraphics[width=2.0\columnwidth]{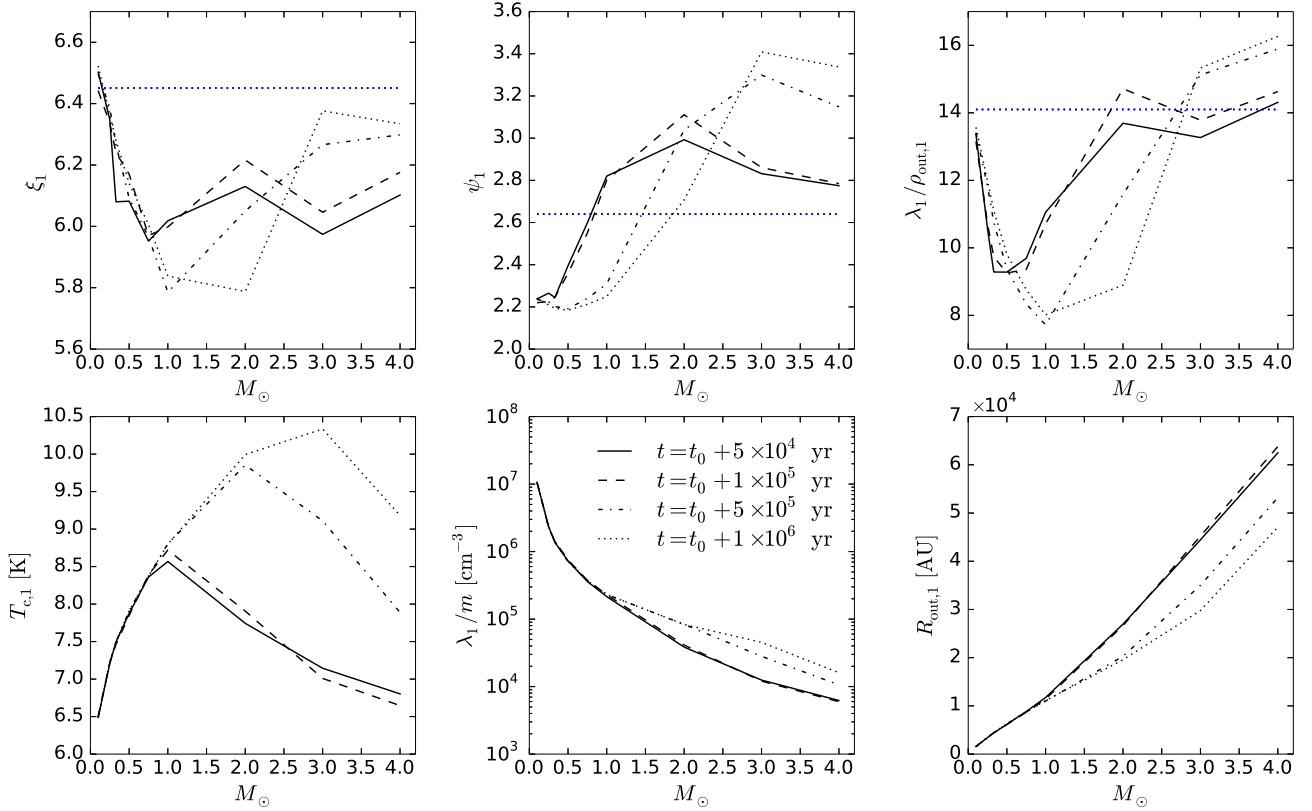}
\caption{Critical non-dimensional radius $\xi_1$ ({\sl upper left-hand} panel), $\psi(\xi_1) = \psi_1$ ({\sl upper middle} panel), density contrast $\lambda_1 / \rho_{\rm out,1} = (T_{\rm out,1} / T_{\rm c,1}) e^{\psi_1}$ ({\sl upper right-hand} panel), central temperature $T_{\rm c,1}\,=\,4\pi Gm\beta_1 / k$ ({\sl lower left}), central density $\lambda_1 / m$ ({\sl lower middle}), and the outer core radius $R_{\rm out,1} = \beta^{1/2} \lambda^{-1/2} \xi_1$ ({\sl lower right}) of each critical core configuration as functions of core mass. The profiles are plotted at four different time steps, given in the legend in the bottom middle panel. The blue dotted lines in the upper panels represent the corresponding critical values for the (isothermal) Bonnor-Ebert sphere ($\xi_0 = 6.45$, $\psi_0 = 2.64$, and $\lambda_0 / \rho_{\rm out,0} = 14.1$).}
\label{fig:stab}
\end{figure*}

We present in Fig.\,\ref{fig:stab} the critical non-dimensional radius $\xi_1$ and other related quantities (see below) as functions of core mass and chemical time, as derived from Eq.\,(\ref{stability}). The values of $\xi_1$ derived here are consistently lower than in Paper~I for the OH94 model (their Fig.\,4). However, the $\xi_1$ values predicted by the two works agree to within $10\,\%$ regardless of core mass. Hence we derive similar stability for (deeply embedded) critical MBESs regardless of whether we assume $T_{\rm dust} = T_{\rm gas}$ or $T_{\rm dust} \neq T_{\rm gas}$.

Evidently, there is marked variation in $\xi_1$ both as a function of core mass and as a function of chemical time. A clear decreasing trend in $\xi_1$ is seen from the lowest masses ($\sim 0.1\,M_{\odot}$) up to about $0.75\,M_{\odot}$. $\xi_1$ coincides with the isothermal critical radius at $\sim 0.25\,M_{\odot}$. For masses above $0.75\,M_{\odot}$, $\xi_1$ is, at early times, constant to an accuracy of $\sim 0.15$ regardless of core mass. However, at late times the value of $\xi_1$ depends on the core mass. In the range $0.75\,M_{\odot} - 2.5\,M_{\odot}$, $\xi_1$ decreases as a function of chemical time while for $M \gtrsim 2.5\,M_{\odot}$ an increase with chemical time is evident. For an isothermal BE sphere, the translation from $\xi_{\rm out}$ to physical quantities is straightforward: a larger $\xi_{\rm out}$ translates directly to a higher density contrast between the center and the edge, while the physical radius depends on $\xi_{\rm out}$, the central density, and the isothermal temperature. Here, the conversion between the various parameters is less evident because the temperature is also a function of radius, and the temperature profile affects not only the density contrast (Eq.\,\ref{eqrho}) but also the solution to the Lane-Emden equation and the determination of the central density (Eqs.\,\ref{eqMLE}~and~\ref{eqBEmass}). To understand the significance of the temporal changes in $\xi_1$, we have to consider the associated changes of $\lambda_1$, $\beta_1$ (i.e., $T_{\rm c,1}$), and $\psi_1$.

The central density of a critically stable non-isothermal sphere, $\lambda_1$, decreases monotonously with increasing mass like in the isothermal case. Chemical evolution starts to modify $\lambda_1$ above $0.75\,M_{\odot}$ where it increases with chemical time. The central temperature $T_{\rm c}$ first increases as a function of mass, reaches a maximum, and then decreases again towards the largest masses. Like $\lambda_1$, $T_{\rm c}$ shows hardly any temporal variation below $0.75\,M_{\odot}$. Beyond this point, however, $T_{\rm c}$ increases strongly with chemical time and the maximum shifts towards larger masses. Finally, $\psi_1$ decreases with chemical time for masses below $\sim 2.5\,M_{\odot}$, but increases for the higher masses.

The approximate constancy of $\xi_1$, $\lambda_1$ and $\beta_1$ over chemical time at the lowest masses ($< 0.75\,M_{\odot}$) implies that the outer radii of critically stable low-mass cores are similar regardless of how long the gas has been chemically processed. For higher core masses, the physical size of a critical core always decreases with chemical time, but the effect is much more prominent for the highest masses. This is a consequence of the increasing average temperature caused by chemical evolution; hotter cores are better able to withstand gravity and external pressure, and thus higher densities are required for the cores to collapse.

\begin{figure*}
\centering
\includegraphics[width=2.0\columnwidth]{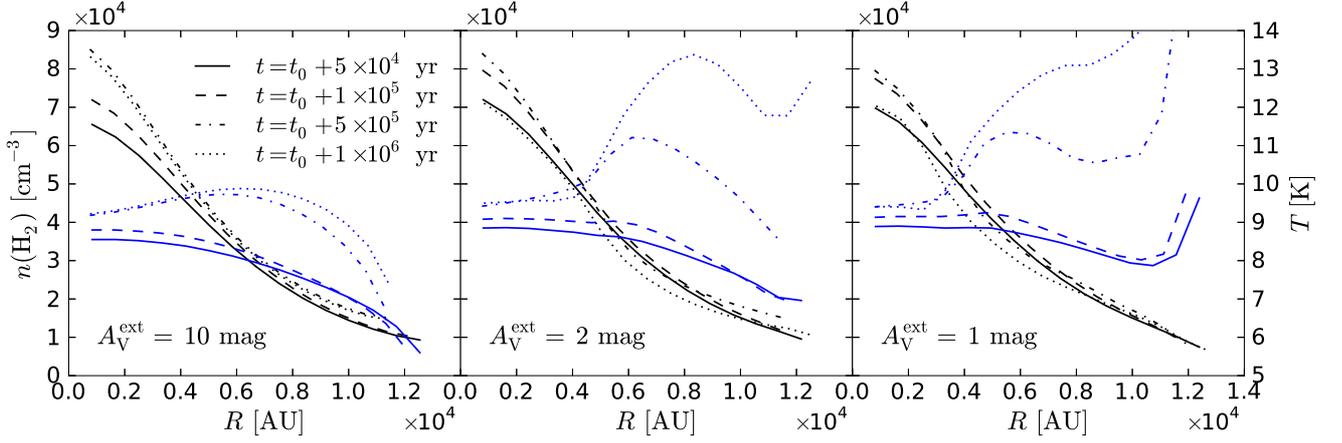}
\caption{Density (black lines, scale on left y-axis) and gas temperature (blue lines, scale on right y-axis) profiles of an MBES with $M = 1\,M_{\odot}$, $\xi_{\rm out} = 5$ as functions of core radius. The profiles are plotted for four different time steps, labeled in the left panel. {\sl Left panel:} External $A_{\rm V} = 10$~mag. {\sl Middle panel:} External $A_{\rm V} = 2$~mag. {\sl Right panel:} External $A_{\rm V} = 1$~mag.
}
\label{fig:avcomp}
\end{figure*}

For critically stable low-mass MBESs, the density constrast is always lower than that of a critical BES ($\sim 14$). For $M~<~0.75\,M_{\odot}$, the density contrast increases with chemical time owing to changes in the temperature profile; the temperature at the edge of the cores increases with CO depletion (see also Sect.\,\ref{ss:ce}; note that the central temperature is not affected owing to the efficient gas-grain thermal coupling at the core center) and the consequent increase in the $T_{\rm out} / T_{\rm c}$ ratio overcompensates the decrease of $e^{\psi}$ due to the decrease of $\psi$ as a function of chemical time. For $M > 0.75\,M_{\odot}$, the density contrast largely follows the changes in $\psi$. For $M > 2.5\,M_{\odot}$, the density contrast can slightly exceed that of the critical BES ($\sim 14$).

The described tendencies are controlled by three processes: 1) thermal coupling between gas and dust, 2) molecular line cooling, and 3) depletion of molecules onto dust grains. The gas-grain thermal coupling is efficient at densities above $n({\rm H_2})~\sim~10^5\,\rm cm^{-3}$. Therefore the lowest-mass stable cores which have the highest (central and average) densities are practically unaffected by changes in the chemical composition of the gas, because the temperature profile is determined by the interaction between the dust and the interstellar radiation field. The same is true for the centers of slightly more massive cores up to about  $1\,M_{\odot}$. However, their outer parts are cooled efficiently by molecular line emission, until molecules, in particular CO, start to freeze out. With the diminished cooling, the gas temperature rises.

The most massive MBESs have low central densities and the temperature is determined by dust heating and molecular line cooling throughout. The temperature increases strongly in their central parts owing to CO depletion. For each time step, there is a local maximum in the central temperature (as a function of core mass) which shifts toward higher core masses for longer chemical times. This is because higher-mass cores have increasingly lower average densities and hence the gas-grain thermal coupling is progressively weaker, allowing the temperature to increase further with CO depletion. However, the depletion timescales are very long in the outer layers where the densities are low, and line cooling can operate there also at late chemical times. The steep temperature gradient results in a large density contrast between the center and edge for the most massive cores ($\sim 16$ for the critical cores).

Given the relatively straightforward temporal changes in $T_{\rm c}$ and $\lambda_1$, we can deduce that the function $\psi_1$ is mainly responsible for the mass-dependent behavior of $\xi_1$ as a function of chemical time. The changes in the cooling efficiency alter the radial density distribution described by $\psi$, which leads either to a decrease or an increase of $\xi_1$ depending on the core mass.

\section{Discussion}\label{s:discussion}

\subsection{The effect of temperature on the results}\label{ss:temp_eff}

We have presented the critical radii of MBESs that are deeply embedded in a parent molecular cloud, corresponding to a high external visual extinction $A_{\rm V}^{\rm ext} = 10$\,mag. This choice facilitates comparison with the results of Paper~I where the same assumption was made. However, as pointed out in Sect.\,\ref{s:results}, the determination of the temperature may have a strong impact on our results because the solution to Eq.\,(\ref{eqMLE}) is not unique. To study how our results might vary with different initial assumptions, we have calculated the properties of a $1.0\,M_{\odot}$ core with two low values of $A_{\rm V}^{\rm ext}$ (2 mag and 1 mag).

Figure \ref{fig:avcomp} presents the density and temperature profiles of a $1.0\,M_{\odot}$ core with (arbitrarily chosen) $\xi_{\rm out} = 5$ at different time steps, assuming $A_{\rm V}^{\rm ext} = 10$, 2 or 1~mag. The gas temperatures are similar at the core center regardless of the choice of external $\rm A_{\rm V}$, because of the high density: even for $A_{\rm V}^{\rm ext} = 1$, the visual extinction at the core center is $\sim 10$\,mag (we assume $\rm n(\rm H_2) / A_{\rm V} = 9.4\times10^{20}$; \citealt{Bohlin78}). However, there are significant differences between the gas temperature profiles outside the core center depending on the choice of external $A_{\rm V}$. This is because a decrease of visual extinction increases the heating by the photoelectric effect in the core. Although the photodissociation rate of the major coolant molecule CO is strongly increased as well, the net effect on core cooling is not large because in these circumstances atomic carbon takes over as the main coolant species. The gas temperature at the edge of the core for $A_{\rm V}^{\rm ext}= 1$ and $t = 10^6$\,yr is $\sim 26\,\rm K$.

\begin{figure}
\centering
\includegraphics[width=\columnwidth]{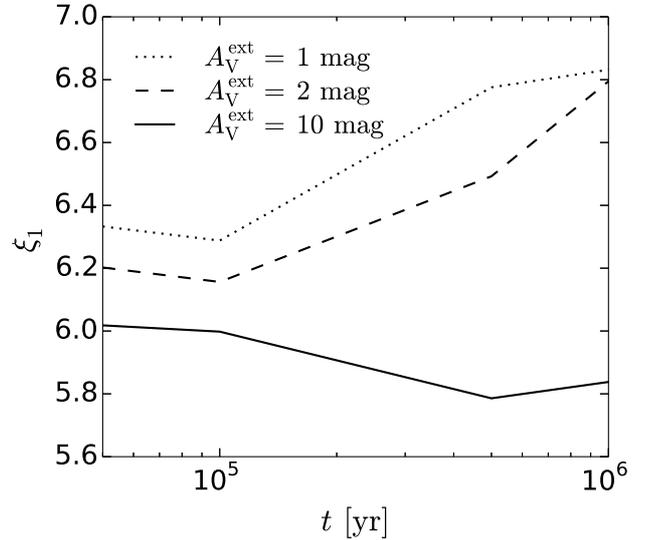}
\caption{Critical non-dimensional radius $\xi_1$ ({\sl upper left} panel) of the $1.0\,M_{\odot}$ MBES as a function of chemical time for different values of $A_{\rm V}^{\rm ext}$, indicated in the figure.}
\label{fig:stab_lowav}
\end{figure}

\begin{figure*}
\centering
\includegraphics[width=2.0\columnwidth]{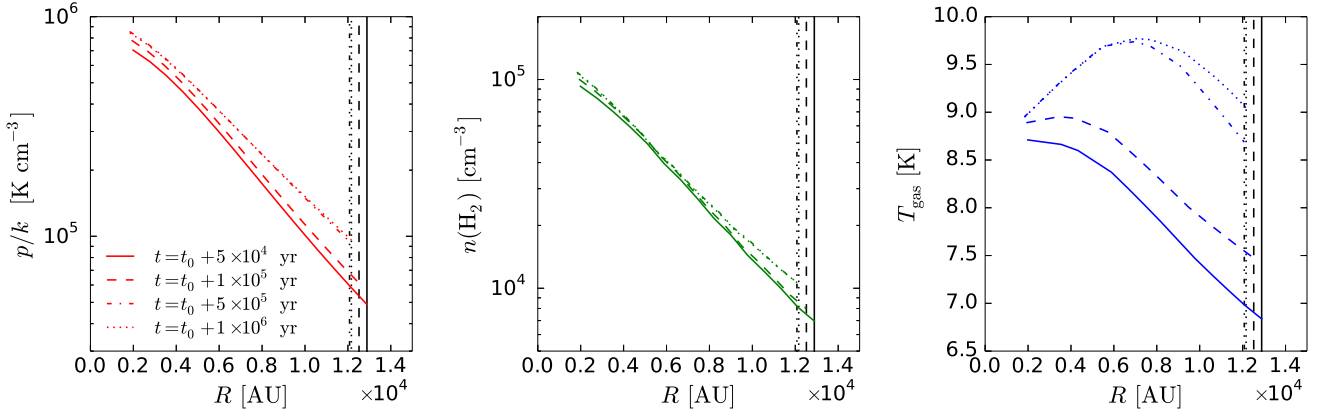}
\caption{Thermal pressure ({\sl left-hand panel}), density ({\sl middle panel}) and gas temperature ({\sl right-hand panel}) of an $M = 1.0\,M_{\odot}$ MBES with $\xi_{\rm out} = 6$ as functions of distance from the core center at four different time steps, labeled in the left-hand panel. The vertical lines in each panel represent the outer radius of the core at the different time steps.
}
\label{fig:ce}
\end{figure*}

Evidently, the central densities are similar in all cases, even though the determination of the central density depends also on the temperature and $\psi$ profiles, and not only on the central temperature (Eq.\,\ref{eqBEmass}). The density profiles are also similar for chemical times up to $t = t_0 + 5 \times 10^5$\,yr. For $t = t_0 + 1 \times 10^6$\,yr at low $A_{\rm V}^{\rm ext}$, the strongly increased gas temperature changes the mass distribution in the core so that the slope of the density profile becomes somewhat steeper.

The differences in temperature depending on the choice of $A_{\rm V}^{\rm ext}$ translate to very different solutions to Eq.\,(\ref{eqMLE}), and consequently the non-dimensional radius of the critical configuration is different in all cases. Figure~\ref{fig:stab_lowav} shows the change of the critical non-dimensional radius $\xi_1$ of the $1.0\,M_{\odot}$ core as a function of chemical time, assuming $A_{\rm V}^{\rm ext} = 10$, 2 or 1~mag. Evidently, the critical radius increases as $A_{\rm V}^{\rm ext}$ decreases. Also, for low $A_{\rm V}^{\rm ext}$, the critical radius tends to increase as a function of chemical time, whereas for $A_{\rm V}^{\rm ext} = 10$\,mag, there is a slight decreasing trend.

We find that the $\xi_1$ values are within $\sim 15\,\%$ of each other regardless of the choice of $A_{\rm V}^{\rm ext}$, when $M = 1.0\,M_{\odot}$. For lower core masses, we expect the results to be closer together because of the high average densities, which lead to significant temperature differences only at the very edge of the cores. However, for $M > 1.0\,M_{\odot}$, the average density is small and hence the $A_{\rm V}$ gradient through the core is shallow, and we expect temperature effects to be more prominent. The increase of $\xi_1$ with decreasing $A_{\rm V}$ implies that stable isolated cores can have a larger density constrast (between the core center and the edge) than cores embedded in molecular clouds.

We note that in Paper~I, we found similar values for $\xi_1$ regardless of the dust model, although the dust temperature given by the model of \citet{LD01} is lower than that given by the model of \citet{OH94} (see Paper~I). In the present paper, we have not considered the \citet{LD01} dust model, as we do not expect our results to be strongly dependent on the dust model. This is motivated on the one hand by the results of Paper~I and on the other hand by the fact that $\xi_1$ does not change significantly even if the gas temperature profile is radically different (Fig.\,\ref{fig:stab_lowav}).

\subsection{Core evolution} \label{ss:ce}

In the stability calculations presented in this paper, we considered a series of model cores defined by the mass and the non-dimensional radius, so that we could derive values for the critical non-dimensional radius according to an analytical formula (Eq.\,\ref{stability}). However, the choice of the parameters used to define and MBES is free and when constructing a model core one could, instead of the non-dimensional radius, set for example the external pressure or the (dimensional) outer radius. In a realistic scenario, one can expect all of these quantities to change as the core and its surroundings evolve; regardless of how the two required parameters are chosen, implicit assumptions are made simultaneously on the properties of the medium outside the core.

If we consider a fixed non-dimensional radius like earlier in this paper, the properties of the core are limited by the external thermal pressure exerted on it. This can be understood by studying the evolution of the medium at the core boundary. In Fig.\,\ref{fig:ce}, we plot the thermal pressure, density and gas temperature profiles of an MBES with $M = 1.0\,M_{\odot}$ and $\xi_{\rm out} = 6$. The $\xi$ grid has been converted to physical radius $R$ using Eq.\,(\ref{eqr}), and vertical lines marking $\xi_{\rm out} = 6$ have been inserted at each timestep to help estimate the change in $R$. The gas temperature increases as a function of time owing to CO depletion. For late chemical times with increasing thermal pressures, the configuration defined by $\xi_{\rm out} = 6$ decreases in size (and increases in average density). We note that the hydrodynamical models of \citet{Keto05} predict that MBESs can exhibit oscillatory behavior. Expansion as a function of time is not achieved in our models because the gas temperature is nearly always increasing, leading to more efficient compression of the cores at long chemical timescales.

The changes in physical radius, density profile etc. at the different time steps considered here do not necessarily represent evolutionary tracks for the cores, because in a realistic scenario neither the non-dimensional radius nor the mass of a core are constant. However, the present study demonstrates that the conditions for the core stability change with chemical evolution. For example, in the case of the $1\,M_{\odot}$ MBES, we expect a core configuration with $\xi_{\rm out} \sim 6$ to be critically stable if the medium is chemically young, while the same configuration would be unstable if associated with chemically old gas. Nevertheless, when a core has been found, e.g., based on the column density and temperature distributions, to agree with a MBES model, the diagrams presented in Fig.\,\ref{fig:stab} of this paper can be used to estimate its stability.

\section{Conclusions}\label{s:conclusions}

We analyzed the stability of non-isothermal (modified) Bonnor-Ebert spheres with a new model that includes a self-consistent determination of the gas temperature. We compared our results with those of \citet{Sipila11}, where it was assumed that $T_{\rm dust} = T_{\rm gas}$. We found that the critical non-dimensional radius $\xi_1$ changes with the chemical evolution in the core and its surroundings, especially for cores with $M > 1\,M_{\odot}$. This is because of the depletion of the coolant species (mainly CO) onto grain surfaces which raises the gas temperature. Therefore, cores that exist in a chemically young environment are expected to have a different stability condition than those that exist in regions of chemically old gas.

The $\xi_1$ values derived in the present work are slightly lower than in the models of \citet{Sipila11}. However, the results of the two works agree to within $10\,\%$. The bulk of our analysis was carried out for deeply embedded cores with external visual extinction $A_{\rm V}^{\rm ext} = 10$\,mag; test calculations for lower values of $A_{\rm V}^{\rm ext}$ yield higher critical radii. In summary, our results indicate that the stability of the modified Bonnor-Ebert sphere is similar regardless of whether one assumes $T_{\rm dust} = T_{\rm gas}$ or $T_{\rm dust} \neq T_{\rm gas}$ and, by extension, also similar to the stability of the classic isothermal Bonnor-Ebert sphere \citep[see][]{Sipila11} -- when one assumes that the core is deeply embedded in a larger structure.

\begin{acknowledgements}

We thank the anonymous referees for helpful comments which improved the paper. O.S. acknowledges financial support from the European Research Council (ERC; project PALs 320620), from the Academy of Finland grant 250741, and from the Department of Physics of the University of Helsinki.

\end{acknowledgements}

\bibliographystyle{aa}
\bibliography{refs.bib}

\end{document}